# Registration of terahertz irradiation with silicon carbide nanostructures

N.T. Bagraev[1,2], S.A. Kukushkin[1], A.V. Osipov[1], L.E. Klyachkin[2], A.M. Malyarenko[2], V.S. Khromov[1,2]

[1] *Institute of Problems of Mechanical Engineering, Russian Academy of Sciences, 199178 St. Petersburg, Russia*
[2] *Ioffe Institute, 194021 St. Petersburg, Russia*
E-mail: bagraev@mail.ioffe.ru, sergey.a.kukushkin@gmail.com

**Abstract** - The response to external terahertz (THz) irradiation from the silicon carbide nanostructures prepared by the method of substitution of atoms on silicon is investigated. The kinetic dependence of the longitudinal voltage is recorded at room temperature by varying the drain-source current in the device structure performed in a Hall geometry. In the frameworks of proposed model based on the quantum Faraday effect the incident radiation results in the appearance of a generated current in the edge channels with a change in the number of magnetic flux quanta and in the appearance of features in the kinetic dependence of the longitudinal voltage. The generation of intrinsic terahertz irradiation inside the silicon carbide nanostructures is also revealed by the electrically-detected electron paramagnetic resonance (EDEPR) measured the longitudinal voltage as a function of the magnetic field value.

## 1. Introduction

Considerable attention is being paid presently to studies focused on the generation and utilization of terahertz (THz) irradiation that occupies the frequency interval of $0.1-10$ THz (1 THz=$10^{12}$ Hz, which corresponds to a wavelength of $\sim 300$ μm and an energy of $\sim 4.14$ meV). This radiation has a number of characteristics that make it promising for applications in various fields.

Potential biological and medical applications of THz radiation are being studied extensively. Since the energies of vibrational transitions in molecules fall within the terahertz range, THz radiation may be used to identify and examine them (specifically, to study DNA [1] and detect nucleic acids, carbohydrates, proteins, and peptides [2]). Terahertz irradiation is non-ionizing and does not damage biomolecules, since the energy of photons is low relative to the ionization energy that may induce atomic or molecular dissociation [3]. Experiments demonstrate that this radiation affects the permeability of red-cell membranes and may thus facilitate the release of hemoglobin [4]. It has also been reported that THz radiation suppresses the expression of genes associated with various skin diseases [5]. In addition, THz radiation is readily absorbed by water and bodily fluids, which makes it a promising tool for early detection of oncological diseases [1].

Potential telecommunication applications of THz radiation are no less appealing [6]. For example, a THz communication system with a silicon-based photonics integrated circuit supporting wireless non-return-to-zero on-off-keying signal transmission at a rate of 40Gb/s was examined in [7]. Rates of data transmission in excess of 100Gb/s over distances longer than 110m were achieved with a 0.3 THz carrier wave [8].

Nonpolar and nonmetal materials, such as paper, plastic, garments, wood, and ceramics, are transparent for THz radiation. Therefore, in contrast to X-ray radiation, THz radiation may be used in security systems to detect concealed objects without any detriment to human health. It is also worth noting that the processes in materials subjected to THz irradiation (rotational transitions in molecules, lattice vibrations in solids, large-amplitude vibrational motion in organic compounds) may be utilized in detection of explosives and chemical and biological weapons [6].

At first, thermometers were used to detect infrared (IR) and THz radiation. Thermal effects induced by incident radiation form the physical basis for a class of so-called thermal THz radiation detectors. A Golay cell, where heat from an absorber irradiated by incident radiation is transferred to a monoatomic gas (argon or xenon) isolated within a cell with a membrane wall, belongs to this class. The gas expands and induces deformation of the membrane wall with a mirror surface, which is irradiated by a light-emitting diode. Radiation reflected from the mirror wall is incident on a photodiode. The illumination intensity measured by it is proportional to the degree of mirror deformation [9]. In solid-state samples, temperature variations translate into changes in spontaneous polarization of samples made from pyroelectric materials, of which triglycine sulfate (TGS) and deuterated triglycine sulfate (DTGS) are common examples. The absorption of heat induces a change in intensity of the electric field between the opposite faces of a sample. This variation may be detected [9]. While thermal detectors of this kind are suitable for room-temperature operation, they are not sufficiently sensitive to detect low-intensity radiation and are affected by displacement and vibrations (this is especially true for pyroelectric detectors, which are also piezoelectric [9]). Owing to the freezing of carriers, the resistance increases significantly in semiconductor bolometers cooled to low temperatures. Following absorption of THz quanta, carriers enter the conduction band, thus inducing a detectable resistance reduction [9]. Mercury cadmium telluride (MCT) is one of the commonly used materials for such devices. Although these detectors feature a fast response and high sensitivity, their application is limited by the need for cryogenic cooling [9]. Several attempts at circumventing this limitation with the use of device structures based on HTSC materials have been made [10], but technological difficulties associated with the production of devices with reproducible characteristics hindered the progress in this research.

Terahertz irradiation may also interact with carriers in a detector, which is then said to belong to the class of so-called electron detectors. Effects underlying their operation may be related to the collective behavior of carriers (interaction of THz radiation with plasma waves in the channel of field-effect transistors, which may be characterized using a hydrodynamic analogy [11]) or to the interaction of photons with individual carriers (e.g., the passage of a carrier through a potential barrier in a Schottky diode). Such detectors usually operate at room temperature and are compact, thus allowing one to construct array structures. The narrow-band nature of defection is one of their drawbacks; in the case of transistor type detectors, this necessitates the use of several antenna structures [9].

Terahertz irradiation receivers may be regarded as devices that are either separate from the source or are associated closely with it. Specifically, the technique of self-mixing involves feeding a part of radiation emitted by a laser back into the cavity and modifying the laser operation parameters. The amplitude and the phase of this reradiated signal may be measured [12]. This interferometry technique implemented with a quantum cascade laser was used to visualize the state of biological tissues [13]. It was demonstrated in recent studies that semiconductor nanosandwiches based on silicon heavily doped with boron may be used as sources of THz radiation in diagnostics of oncological diseases performed in accordance with the above method [14]. Radiation is generated in these nanosandwiches in edge channels confined by dipole centers that form networks of Josephson junctions [15]. It was found recently that silicon carbide, which provides an opportunity to raise the radiation power, may be used as the base material for a nanosandwich emitter [16]. Therefore, the examination of possibilities of detection of THz radiation at room temperature with the use of a nanosandwich of this kind is a topical issue.

## 2. Experimental methods

Semiconductor samples with single-crystalline SiC films grown on the surface of single-crystalline silicon were used as the basis for a detector. The procedure of their fabrication by coordinated substitution of atoms with the use of a chemical reaction between silicon and carbon monoxide gas was detailed in [17–19]. An in-depth description of the accompanying processes was given in reviews [20,21].

This SiC film growth mechanism differs from the other methods in that the structure of the initial cubic Si lattice persists, thus providing for the growth of the cubic 3$C$-SiC [20–22] polytype. This was verified in electron microscopic studies, which also revealed the lack of lattice misfit dislocations at the 3C-SiC(111)/Si(111) interface; instead, stacking faults with interlayers of hexagonal phases are present at the interphase boundary [23]. The term "coordinated" implies that the processes of removal of a Si atom from the lattice and introduction of a C atom into the vacant position in reaction

$$2\text{Si (crystal)} + \text{CO (gas)} = \text{SiC (crystal)} + \text{SiO (gas)} \uparrow \tag{1}$$

are concurrent [22]. With the mentioned lack of misfit dislocations, epitaxy of silicon carbide films on silicon due to the coordinated substitution of one half of Si atoms with C atoms ensures high crystalline perfection of SiC films [20–23].

The synthesis of silicon carbide in reaction (1) is a two-stage process. Silicon vacancy–interstitial carbon atom complexes form first. Carbon atoms then shift toward silicon vacancies with the formation of silicon carbide. Activated complexes transform into silicon carbide, and free vacancies assemble in pores below the SiC layer. The end result is the formation of a silicon carbide film partially suspended above pores in silicon. This is the reason why films formed this way are free from elastic strain [17–23]. In contrast to traditional growth methods, the film orientation is set not just by the substrate surface, but by the "old" crystal structure of the initial silicon matrix.

The emergence of a layer with unusual optical and electrophysical properties and a thickness on the order of several nanometers at the SiC/Si interface is an important feature of this technique of synthesis of silicon carbide by coordinated substitution of atoms. Its formation is induced by the process of contraction of the initial Si lattice with a parameter of 0.543 nm, which "collapses" to a cubic SiC lattice with a parameter of 0.435 nm, at the final stage of transformation of silicon into silicon carbide. This process occurs in the substrate plane [20,23]. Silicon carbide detached from the silicon matrix subjects it to anomalously strong compression (exceeding 100GPain magnitude). It would be impossible to produce SiC with a structure ordered so tightly under such pressures if every fifth lattice cell of silicon carbide did not align accurately with every fourth cell of silicon. As a result
of material contraction, every fifth chemical bond of SiC is positioned in a coordinated fashion with every fourth bond of Si. The other bonds either get disrupted, thus producing vacancies and pores, or are subjected to compression, which alters the structure of surface bands of silicon carbide adjacent to Si and leads to its transformation into a "semimetal". This effect has been observed for the first time in a recent study performed using the spectral ellipsometry technique in the 0.5−9.3 eV range of photon energies [24].

It follows from the results of quantum-chemical calculations [24] that, in the process of dislocation-free matching of SiC and Si lattices that differ by 20%, a SiC film with its Si surface facing the substrate attracts one of the 16 Si atoms in the proximal double layer of substrate atoms. Out of 25 Si atoms, 22 form chemical bonds with Si substrate atoms, while the remaining three (i.e., 12%) do not form bonds, since they are located too far (> 3 Å) from substrate atoms. These are the Si atoms in SiC the $p$ electrons of which produce the primary contribution to the narrow and well-pronounced 3$C$-SiC(111)/Si(111) peak of the electron state density located in the vicinity of the Fermi energy. In other words, the 3C-SiC(111)/Si(111) interface should exhibit unusual electrophysical properties; specifically, it should be a fine electric conductor.

This interface also exhibits certain unusual magnetic properties. A sample with a silicon carbide film grown on (110) silicon was examined in [25]. Following the formation of this film, doping with boron in the conditions of non-equilibrium gas-phase diffusion was performed; the technological parameters for this sample (F5) are given in [25]. The observed "dia−para-hysteresis" of the magnetic susceptibility amounts to an experimental demonstration of the Meissner−Ochsenfeld effect. Oscillations of the magnetic susceptibility, which signify the fulfillment of conditions of quantum interference in the vicinity of microdefects in the sample plane, were also noted. The examination of these oscillations allowed us to identify the de Haas−van Alphen (DHVA) and Aharonov−Bohm (AB) effects, which are associated with quantization of the moment and the magnetic flux at room temperature, respectively, due to the effective suppression of the electron–electron interaction at high temperatures. This is made possible by the formation of dipole boron centers with a negative correlation energy that confine the edge channels of the studied structure and, owing to their interaction with carriers, govern the characteristics of DHVA and AB oscillations [26]. It should be noted that edge channels may be formed not only by dipole boron centers, but also by dipole centers of the "silicon vacancy–interstitial carbon atom" type, which are always present in SiC/Si structures grown by coordinated substitution of atoms on the (111) and (110) silicon substrate surfaces [22]. Thus, the observation of quantum interference in edge channels is interrelated with the above-mentioned "dia−para-hysteresis" of the magnetic susceptibility.

Surface contacts were formed in the Hall geometry (see Fig. 1) to carry out experiments with current flowing through the detector sample based on silicon carbide.

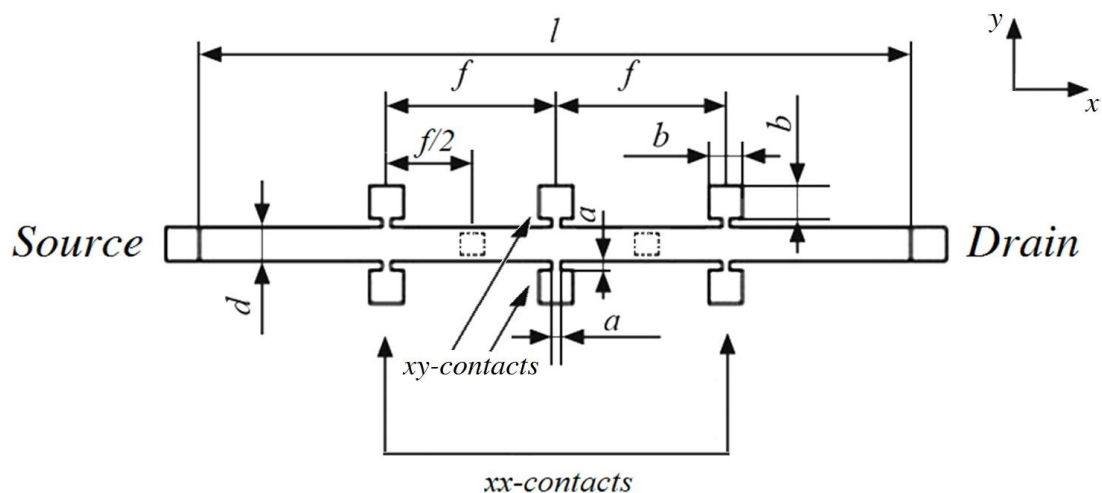

**Figure 1.** Hall geometry of contacts on the surface of the studied SiC-based structure. Parameters ($\mu$m): $a = 50$, $b = 200$, $d = 200$, $l = 4200$, $f = 1000$. Dashed contours denote the positions of vertical contacts $b \times b$ in size formed above the structure surface.

The measurement diagram presented in Fig. 2, *a* was implemented in these experiments. The studied detector sample based on silicon carbide was positioned at distance *s* from the THz radiation emitter. The detector was housed in a protective metallic box with an aperture. A Tydex LPF14.3 THz filter blocking radiation with frequencies > 14.3 THz was mounted in front of the aperture. A nanosandwich based on silicon heavily doped with boron was used as the THz radiation emitter. It contained an ultra-narrow silicon quantum well (2 nm in width) confined by two $\delta$-barriers containing dipole boron centers with a negative correlation energy [27]. This emitter has the geometry of a Hall bridge (similar to the one presented in Fig. 1) and is characterized by the following parameters ($\mu$m): $a = 50$, $b = 200$, $d = 200$, $l = 4720$, and $f = 1000$. The generation of THz radiation is induced by the passage of longitudinal source−drain current $I_{ds\,(emi)}$ (Fig. 2, *b*) of the milliampere range through the emitter [27]. Both samples were housed in aluminum enclosures with an aperture and were positioned with their plane surfaces facing each other.

Longitudinal voltage $U_{ds}$ of the detector was measured in the experiment as a function of time with longitudinal source−drain current $I_{ds} = 1.5~\mu A$ flowing through the sample (Fig. 2, c). These measurements were performed at room temperature. A total of 580−1500 measurements with an overall duration of 245−634 s, respectively, were carried out in a single run. Terahertz generation was lacking in the first 200 measurements; the detector was irradiated (with longitudinal current $I_{ds\,(emi)} = 30$ mA applied to the emitter) in all measurements starting from the 201st one.

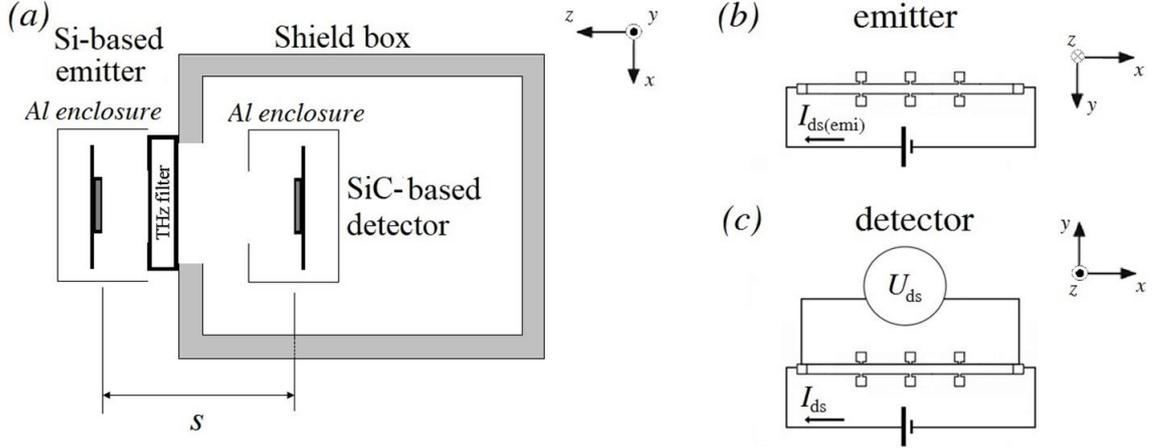

**Figure 2.** *a* — Mutual positioning of the Si-based radiation emitter and the SiC-based THz radiation detector; *s* — distance between the samples. *b* — Diagram of passage of longitudinal source−drain current $I_{ds(emi)}$ through the emitter sample. *c* — Diagram of passage of longitudinal source−drain current $I_{ds}$ through the detector sample and measurement of longitudinal detector voltage $U_{ds}$.

In view of this, quantity

$$\Delta U_{ds} = U_{ds} - \langle U_{ds}(1-200) \rangle, \qquad (2)$$

which is equal to the difference between the experimental value of $U_{ds}$ and its mean value determined in the first 200 measurements, was used for analysis.

## 3. Results and discussion

The results of measurements are presented in Fig. 3, where the vertical line at Counts= 200 marks the initial moment of generation when the emitter current is switched on. The characteristic step features of the kinetic $1U_{ds}(t)$ dependence are reproduced at different distances *s* = 42 (Fig. 3, *a*), 36 (Fig. 3, *b*), and 30 mm (Fig. 3, *c*). A feasible model relating these $\Delta U_{ds}(t)$ features to the detected radiation intensity via the Faraday formula is proposed below. The discussed experiment is specific in that the trapping of single magnetic flux quanta in quantum interference regions is possible. This is manifested in the discovered magnetic susceptibility oscillations with their period depending on the size of the quantum interference region [25]. Thus, it is fair to say that we observe the quantum Faraday effect induced by the trapping of single magnetic flux quanta (magnetic field lines) in quantum interference regions.

Characteristic size *L* of the region of quantum carrier interference may be estimated based on the results of measurements in a magnetic field. According to the magnetic-field dependences of longitudinal voltage $U_{xx}$ of the studied sample, the region of interference of a single carrier in the edge channel is 134 $\mu$m×1.54 nm in size [16]. Since the electron–electron interaction may be suppressed strongly under high pressures on the order of several hundred GPa at the silicon substrate−silicon carbide interface (see above), the formation of interference regions containing a pair of carriers with $L_1 = 268~\mu m$ (i.e., twice as big) is possible. The characteristic size of regions allowing for interference of carrier pairs may be estimated based on the measurement data on magnetic susceptibility oscillations:

$$R^2 = \frac{\Phi_0}{\pi \Delta B}, \quad (3)$$

where $\Phi_0 = h/2e$ is the magnetic flux quantum and $\Delta B$ is the period of magnetic susceptibility oscillations. The determined periods of 13 and 300Oe [25] translate into $R$ values of 0.712 and 0.148 $\mu$m, respectively. Characteristic size $L = 2R$ of the interference region is then $L_2 = 1.424$ $\mu$m and $L_3 = 0.296$ $\mu$m.

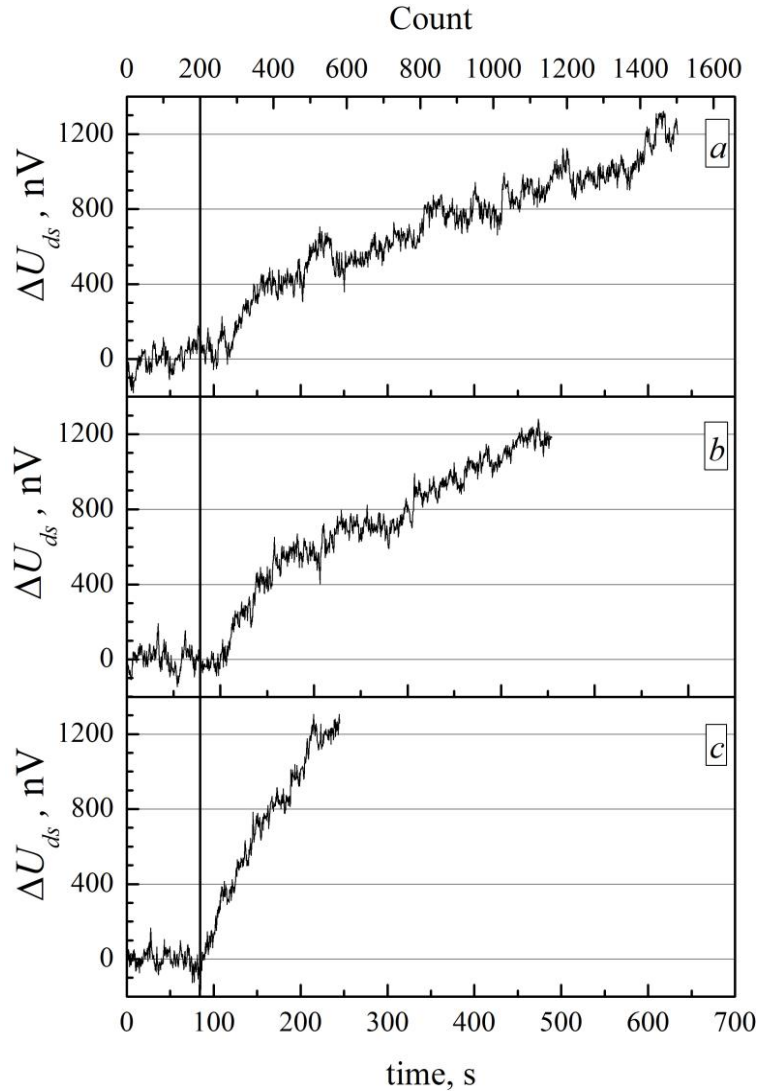

**Figure 3.** Dependence $\Delta U_{ds}$ ($t$) for the SiC-based detector at distances $s$ = 42 (*a*), 36 (*b*), 30mm (*c*). Step features at 400 and 1200 nV correspond to the detection of frequencies of 2.745 THz and 9.096GHz. The vertical line marks the initial moment of generation when emitter current $I_{ds(emi)}$ = 30mA is switched on. Temperature $T$ = 300K, $I_{ds}$ = 1.5 $\mu$A.

Knowing the characteristic sizes of regions of quantum interference of carrier pairs, one may relate the observed $\Delta U_{ds}$ ($t$) features to the incident radiation frequency using the Faraday formula:

$$I_{gen} = \frac{\Delta E}{\Delta \Phi} = \frac{h\nu}{\Phi}, \quad (4)$$

where $I_{gen}$ is the generation current produced after the introduction of additional energy $\Delta E$ into the system in the presence of magnetic flux variation $\Delta \Phi$. Relation $\Delta \Phi = \Phi_0 = h/2e$ ($\Delta \Phi = \Delta B \cdot S$, where

$\Delta \Phi B$ is the field variation upon trapping of a single magnetic flux quantum $\Delta\Phi_0$ in a quantum interference region with area $S$) holds true for a carrier pair in the context of trapping of isolated field lines in a quantum interference region. Therefore, he generation current induced by radiation incident on the detector is related to the corresponding voltage via conductance quantum $G_0 = 2e^2/h$ in the following way: $I_{gen} = 2G_0U$. The end result is the following expression that may be used to estimate the incident radiation frequency:

$$\nu = N \cdot G_0 \Delta U_{ds}/e, \qquad (5)$$

where $N = L_0/L_i$ is the number of quantum interference regions with characteristic size $L_{1,2,3}$ connected in parallel within distance $L_0$ between the measurement contacts. In the case of $ds$ contacts, $L_0 = l = 4200\ \mu m$. The feature at $\Delta U_{ds} = 400$ nV characterizes the contribution of regions with $L_3 = 0.296\ \mu m$ to the generation current and corresponds to $\nu = 2.745$ THz. The feature at $\Delta U_{ds} = 1200$ nV characterizes the contribution of regions with $L_1 = 268\ \mu m$ to the generation current and corresponds to $\nu = 9.096$ GHz.

In order to resolve finer features, one may perform measurements in the same geometry, but using $xx$ contacts as measurement ones (see Fig. 1). The results of such measurements are presented in Fig. 4. In this scenario, $L_0 = 2 f = 2000\ \mu m$; the feature at $\Delta U_{ds} = 220$ nV characterizes the contribution of regions with $L_2 = 1.424\ \mu m$ to the generation current and corresponds to $\nu = 0.15$ THz. The obtained values agree well with the key frequencies of the emitter sample (2.8 and 0.12 THz and 9.3GHz [27]).

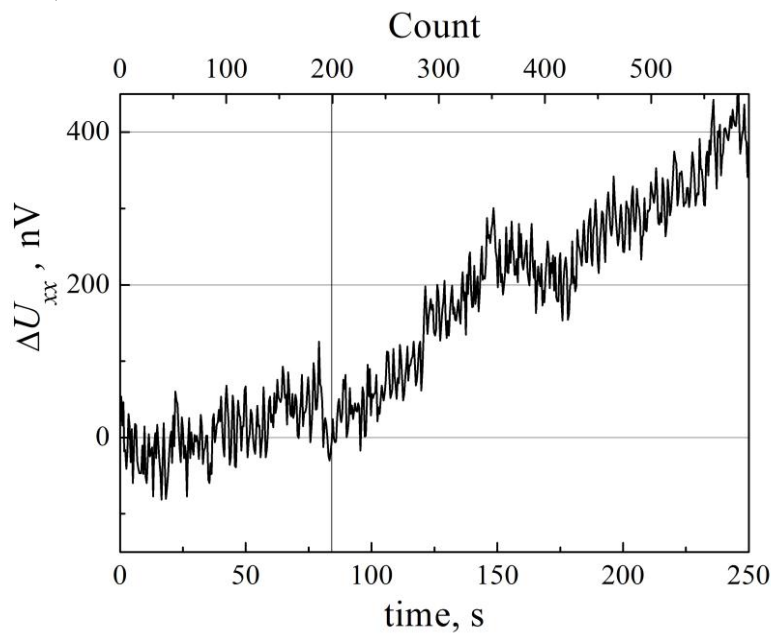

**Figure 4.** Dependence $\Delta U_{xx}(t)$ for the SiC-based detector at $s = 30$mm. The step feature at 220 nV corresponds to the detection of a frequency of 0.15 THz. The vertical line marks the initial moment of generation when emitter current $I_{ds(emi)} = 30$mA is switched on. Temperature $T = 300$K, $I_{ds} = 1.5\ \mu A$.

It should be noted that a component associated with the intrinsic generation of THz radiation due to the passage of longitudinal current along edge channels may be present in the detector response to external THz irradiation. The energy variation in formula (4) is then defined by the load resistance in the quantum interference region. In other words, electrical characteristics in the quantum interference region govern the frequency of generation, which may be estimated by recording the electrically detected electron paramagnetic resonance (EDEPR) spectrum in measurements of magnetic-field dependences of the longitudinal voltage [28]. When nanoampere-range source−drain current flows through the sample with the Hall geometry, microwave generation is observed in the edge channel if embedded microcavities are present in it [28]. In this context, the EDEPR spectrum of point centers localized within edge channels is obtained by scanning over the magnetic field.

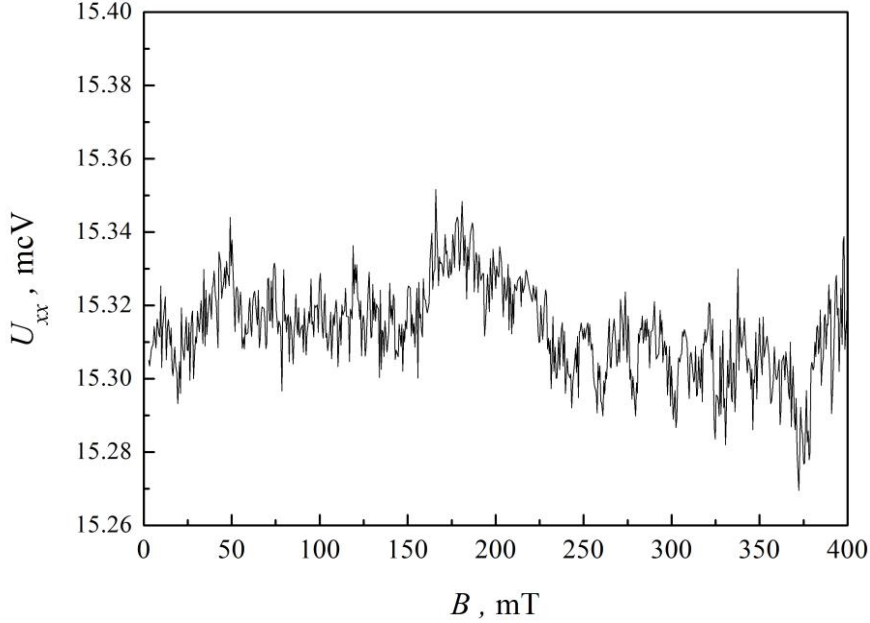

**Figure 5.** EDEPR spectrum of the detector sample recorded by measuring longitudinal voltage $U_{xx}$. The feature at 162.5mT corresponds to a frequency of 3.4 THz. Temperature $T = 300$K, $I_{ds} = 10$ nA.

The results of measurements are presented in Fig. 5, where the room-temperature dependence of voltage $U_{xx}$ at the studied detector sample on magnetic field $B$ applied perpendicularly to the sample plane with $I_{ds} = 10$ nA is shown. It follows from the analysis of the EDEPR spectrum in Fig. 5 that it contains a fragment of the magnetic-field dependence corresponding to the EPR spectrum of a silicon vacancy recorded at a frequency of 9.4GHz (Fig. 6) [29]. This verifies the presence of microcavities supporting the generation and detection of radiation with centimeter wavelengths. In the studied sample, the silicon substrate extending throughout its length apparently acts as a microcavity with geometric length $l_0$: if one assumes the refraction index of silicon to be equal to 3.42 (NSM database [30]), the $l_0 = c/2\nu n$ condition for frequency $\nu = 9.3$GHz yields $l0 = 4.74$ mm, which agrees closely with the value of sample length $l + 2b$ determined with account for the size of contact pads (see Fig. 1).

If the formation of several types of microcavities is feasible and a number of quantum interference regions of different size for radiation generation are present, multifrequency EDEPR may be implemented instead of the single-frequency variant. The magnetic-field dependence recorded in such measurements reproduces the variety of generated frequencies in quantum interference regions. Specifically, the low-field part of the EDEPR spectrum in Fig. 5, $g = 1500$, corresponds to the generation of radiation with a frequency of 3.4 THz, which was found in the electroluminescence spectrum of the studied sample [16].

Owing to the size difference between the corresponding microcavities, 17.3 $\mu$m (see Fig. 9 in [16]), and the area occupied by a single carrier in the edge channel, 134 $\mu$m×1.54 nm [16], the EDEPR spectrum is split into seven components. The excited states of complexes of silicon vacancies interacting with single carriers are revealed clearly as lines of different polarity in strong and weak magnetic fields. This is indicative of their strong spin polarization. It should be noted that the EDEPR signal of multicomponent vacancy centers involved in the exchange interaction with carriers is detected reliably under the condition that the effective mass of a carrier in the edge channel is small:

$$\hbar\omega_c = 2\pi\nu\hbar = \hbar\frac{e\Delta B}{m^*}, \qquad (6)$$

where $m^*$ is the effective mass of a carrier in the edge channel, $\nu$ is the EDEPR spectrum recording frequency, $\Delta B$ is the FWHM of the EDEPR spectrum, and $e$ is the charge of an electron. The effective mass estimated using the magnetic-field dependence in Fig. 5 at $\Delta B = 6.5$mT and $\nu = 3.4$ THz is $m^* = 5 \cdot 10^{-35}$ kg, which agrees with the measurement data on DHVA oscillations [25]. Thus, the observation of EDEPR is actually feasible at a low value of the effective carrier mass, which corresponds with the transport conditions in edge channels and quasi-one-dimensional structures.

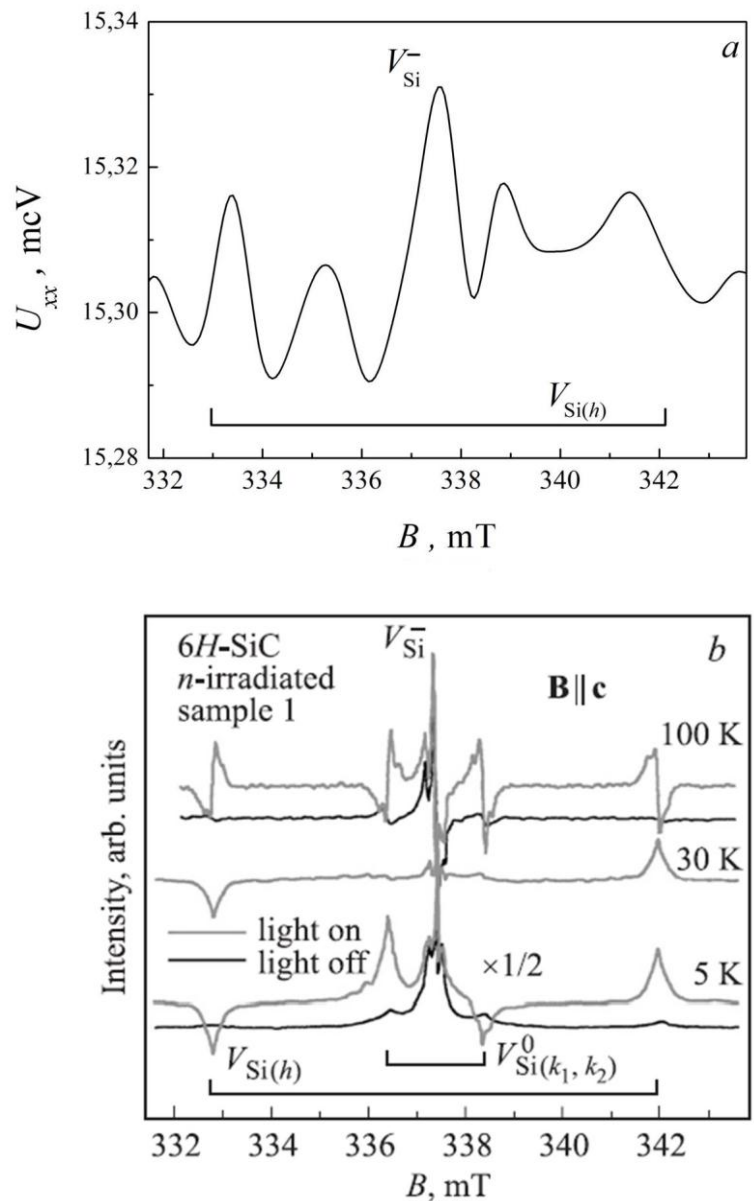

**Figure 6.** *a* — EDEPR spectrum of a silicon vacancy in the detector sample recorded by measuring longitudinal voltage $U_{xx}$ without an external cavity, source, and receiver of microwave radiation; $T = 300$K, $I_- = 10$ nA. *b* — EPR spectrum (*X* band) of a silicon vacancy in 6*H*-SiC (according to [29]).

## 4. Conclusion

Thus, the specific features of response of silicon carbide nanostructures, which were produced by coordinated atomic substitution, to external THz irradiation were revealed and studied. These effects were manifested in room-temperature measurements of the kinetic dependence of longitudinal voltage with a longitudinal source−drain current flowing through the structures of a Hall geometry. The discovered features of the kinetic dependences of longitudinal voltage were examined as manifestations of the quantum Faraday effect emerging in the case of trapping of single magnetic flux quanta by the edge channels of nanostructures. Within the proposed model, THz radiation induces current generation in an edge channel and, consequently, alters the kinetic dependences of longitudinal voltage, which are governed by the geometric parameters of the studied nanostructures.

The technique of EDEPR detection via the measurement of magnetic-field dependences of longitudinal voltage revealed the generation of intrinsic THz radiation in silicon carbide nanostructures with a longitudinal source−drain current flowing through them. It was demonstrated that microcavities embedded into the edge channels of the nanostructure support the generation and detection of THz radiation. EDEPR spectra may be measured reliably at low values of the effective carrier mass in the edge channel of the examined structure.

**Funding**

This work was supported financially by the Russian Science Foundation (grant № 20-12-00193).

**Acknowledgments**

The synthesis of a SiC layer on Si was performed using the equipment of the "Physics, Chemistry, and Mechanics of Chips and Thin Films" unique scientific unit at the Institute of Problems of Mechanical Engineering, Russian Academy of Sciences (St. Petersburg).

**Conflict of interest**
The authors declare that they have no conflict of interest.

**References**
[1] M. Danciu, T. Alexa-Stratulat, C. Stefanescu, G. Dodi, B.I. Tamba, C. Teodor Mihai, G.D. Stanciu, A. Luca, I.A. Spiridon, L.B. Ungureanu, V. Ianole, I. Ciortescu, C. Mihai, G. Stefanescu, I. Chirila, R. Ciobanu, V.L. Drug. Materials (Basel, Switzerland), **12** (9), 1519 (2019).
[2] X. Yang, X. Zhao, K. Yang, Y. Liu, Y. Liu, W. Fu, Y. Luo. Trends Biotechnol., **34**, 810 (2016).
[3] O.P. Cherkasova, D.S. Serdyukov, A.S. Ratushnyak, E.F. Nemova, E.N. Kozlov, Yu.V. Shidlovskii, K.I. Zaytsev, V.V. Tuchin. Opt. Spectrosc., **128** (6), 855 (2020).
[4] S.A. Il'ina, G.F. Bakaushina, V.I. Gaiduk, A.M. Khrapko, N.B. Zinov'eva. Biofizika, **24**, 513 (1979).
[5] L.V. Titova, A.K. Ayesheshim, A. Golubov, R. Rodriguez-Juarez, R. Woycicki, F.A. Hegmann, O. Kovalchuk. Sci. Rep., **3**, 1 (2013).
[6] J. Xie, W. Ye, L. Zhou, X. Guo, X. Zang, L. Chen, Y. Zhu. Nanomaterials, **11**, 1646 (2021).
[7] J. Kim, S.R. Moon, S. Han, S. Yoo, S.H. Cho. Opt. Express, **28**, 23397 (2020).
[8] T. Harter, C. Fullner, J.N. Kemal, S. Ummethala, M. Brosi, E. Br¨undermann, W. Freude, S. Randel, C. Koos. *110-m THz Wireless Transmission at 100 Gbit/s Using a Kramers-Kronig Schottky Barrier Diode Receiver*. In: *Proc. Eur. Conf. on Optical Communication* (*ECOC*), (Roma, Italy, 2018).
[9] R.A. Lewis. J. Phys. D: Appl. Phys., **52**, 433001 (2019).
[10] L. Ozyuzer, A.E. Koshelev, C. Kurter, N. Gopalsami, Q. Li, M. Tachiki, K. Kadowaki, T. Yamamoto, H. Minami, H. Yamaguchi, T. Tachiki, K.E. Gray, W.-K. Kwok, U. Welp. Science, **318**, 1291 (2007).
[11] M. Dyakonov, M. Shur. IEEE Trans. Electron Dev., **43** (3), 380 (1996).
[12] T. Taimre, M. Nikoli´c, K. Bertling, Y.L. Lim, T. Bosch, A.D. Raki´c. Adv. Opt. Photon., **7**, 570 (2015).
[13] Y.L. Lim, T. Taimre, K. Bertling, P. Dean, D. Indjin, A. Valavanis, S.P. Khanna, M. Lachab, H. Schaider, T.W. Prow, H.P. Soyer, S.J. Wilson, E.H. Linfield, A.G. Davies, A.D. Rakić. Biomed. Optics Express, **5** (11), 3981 (2014).
[14] K.B. Taranets, M.A. Fomin, L.E. Klyachkin, A.M. Malyarenko, N.T. Bagraev, A.L. Chernev. J. Appl. Phys., **125**, 225702 (2019).
[15] N.T. Bagraev, L.E. Klyachkin, A.A. Kudryavtsev, A.M. Malyarenko, V.V. Romanov. Semiconductors, **43** (11), 1441 (2009).
[16] N.T. Bagraev, S.A. Kukushkin, A.V. Osipov, L.E. Klyachkin, A.M. Malyarenko, V.S. Khromov. Fiz. Tekh. Poluprovodn., **55** (11), 1027 (2021) (in Russian).

[17] S.A. Kukushkin, A.V. Osipov. Phys. Solid State, **50**, 1238 (2008).
[18] S.A. Kukushkin, A.V. Osipov. Dokl. Phys., **57**, 217 (2012).
[19] S.A. Kukushkin, A.V. Osipov. Mech. Solids, **48** (2), 216 (2013).
[20] S.A. Kukushkin, A.V. Osipov. J. Phys. D: Appl. Phys., **47** (31), 313001 (2014).
[21] S.A. Kukushkin, A.V. Osipov, N.A. Feoktistov. Phys. Solid State, **56**, 1507 (2014).
[22] S.A. Kukushkin, A.V. Osipov. J. Phys. D: Appl. Phys., **50** (46), 464006 (2017).
[23] L.M. Sorokin, N.V. Veselov, M.P. Shcheglov, A.E. Kalmykov, A.A. Sitnikova, N.A. Feoktistov, A.V. Osipov, S.A. Kukushkin. Tech. Phys. Lett., **34** (11), 992 (2008).
[24] S.A. Kukushkin, A.V. Osipov. Tech. Phys. Lett., **46** (11), 1103 (2020).
[25] N.T. Bagraev, S.A. Kukushkin, A.V. Osipov, V.V. Romanov, L.E. Klyachkin, A.M. Malyarenko, V.S. Khromov. Semiconductors, **55** (2), 137 (2021).
[26] N.T. Bagraev, V.Yu. Grigoryev, L.E. Klyachkin, A.M. Malyarenko, V.A. Mashkov, V.V. Romanov. Semiconductors, **50** (8), 1025 (2016).
[27] N.T. Bagraev, L.E. Klyachkin, A.M. Malyarenko, B.A. Novikov. Biotekhnosfera, **5** (41), 55 (2015) (in Russian).
[28] N.T. Bagraev, D.S. Gets, E.N. Kalabukhova, L.E. Klyachkin, A.M. Malyarenko, V.A. Mashkov, D.V. Savchenko, B.D. Shanina. Semiconductors, **48** (11), 1467 (2014).
[29] P.G. Baranov, A.P. Bundakova, A.A. Soltamova, S.B. Orlinskii, I.V. Borovykh, R. Zondervan, R. Verberk, J. Schmidt. Phys. Rev. B, **83**, 125 203 (2011).
[30] http://www.ioffe.ru/SVA/NSM.